\documentclass[12pt]{article}
\usepackage[dvipdfmx]{graphicx}
\usepackage{epstopdf}
\usepackage{bm}
\usepackage{amssymb,amsfonts}
\usepackage[dvipdfmx]{hyperref} 
\usepackage{amsmath}
\usepackage[capitalise]{cleveref}
\usepackage{epsf}
\usepackage{cancel}
\usepackage{cite}

\newcommand{\ave}[1]{\left \langle #1 \right \rangle}
\newcommand{\pdiff}[3]{
\if 1#1   \frac{\partial #2 }{\partial #3 }
\else  \frac{\partial^#1#2 }{\partial #3^#1 } \fi}
\newcommand{\diff}[3]{
\if 1#1  \frac{{\rm d} #2 }{{\rm d} #3 }
\else  \frac{{\rm d}^{#1} #2 }{{\rm d}#3^{#1} } \fi
}         
                          
\newcommand{\dps}[0]{\displaystyle}
\newcommand{\lam}{\lambda}

\newcommand{\ep}{\epsilon}

\newcommand{\ovl}{\overline}
\newcommand{\si}{\sigma}

\newcommand{\eqs}[1]{\begin{equation}\begin{split} #1 \end{split}\end{equation}}

\begin{document}

\begin{titlepage}

\hfill IPMU~15-0103 \\

\begin{flushright}
\end{flushright}

\vskip 1.35cm
\begin{center}

{\large 
{\bf 
Nucleon Electric Dipole Moments \\
in High-Scale Supersymmetric Models}}
\vskip 1.2cm

Junji Hisano$^{a,b,c}$, 
Daiki Kobayashi$^b$,
Wataru Kuramoto$^b$
and
Takumi Kuwahara$^b$\\

\vskip 0.4cm

{\it $^a$Kobayashi-Maskawa Institute for the Origin of Particles and the Universe (KMI),
Nagoya University, Nagoya 464-8602, Japan}\\
{\it $^b$Department of Physics,
Nagoya University, Nagoya 464-8602, Japan}\\
{\it $^c$
Kavli IPMU (WPI), UTIAS, University of Tokyo, Kashiwa, Chiba
 277-8584, Japan}
\date{\today}

\vskip 1.5cm

\begin{abstract} 
  The electric dipole moments (EDMs) of electron and nucleons are
  promising probes of the new physics.  In generic high-scale
  supersymmetric (SUSY) scenarios such as models based on mixture of
  the anomaly and gauge mediations, gluino has an additional
  contribution to the nucleon EDMs.  In this paper, we studied the
  effect of the $CP$-violating gluon Weinberg operator induced by the
  gluino chromoelectric dipole moment in the high-scale SUSY
  scenarios, and we evaluated the nucleon and electron EDMs in the
  scenarios.  We found that in the generic high-scale SUSY models, the
  nucleon EDMs may receive the sizable contribution from the Weinberg
  operator. Thus, it is important to compare the nucleon EDMs with the
  electron one in order to discriminate among the high-scale SUSY
  models.

\end{abstract}

\end{center}
\end{titlepage}

\section{Introduction}
The standard model (SM) is established by the discovery of the 
Higgs boson with mass of 125\,GeV at the LHC run 1
\cite{Aad:2012tfa,Chatrchyan:2012ufa,Aad:2015zhl}.  However, there are
several reasons to expect that the SM must be an effective
theory of a certain full theory; no candidate of the dark matter (DM),
no reason of the gauge anomaly cancellation, and so on.
Supersymmetry (SUSY) is one of the attractive extensions of the SM.
The lightest supersymmetric particle (LSP) is the candidate of the DM.
The gauge coupling unification is improved due to the
additional matters, that is, the SUSY partners of the SM particles.
This unification may imply that the SM gauge groups are embedded in a
larger gauge group, such as $SU(5), SO(10)$, and $E_6$.

However, the LHC run 1 has also reported that there is no signal of
new physics around electroweak (EW) scale.  Besides, the observed
Higgs boson is too heavy in the minimal supersymmetric standard model
(MSSM) if the SUSY particle masses are smaller than $O(1)$~TeV.
It is needed to introduce the large quantum correction to
the Higgs mass or the additional tree-level contribution. Several
extensions of the MSSM are proposed; introduction of additional vector-like
matters \cite{Martin:2009bg}, specific mass spectra (large A-term
or Next-to-MSSM) \cite{Hall:2011aa}, and high-scale SUSY scenarios
\cite{Giudice:2011cg,Ibe:2011aa,Ibe:2012hu}.

In the high-scale SUSY scenarios, sfermions have masses
around $10^{2} ~{\rm TeV}$, while the 
gaugino masses lie around several TeV.  This mass spectrum leads to the
fascinating results: the SUSY flavor and $CP$ problems are eased 
due to heavy sfermions \cite{Gabbiani:1996hi}, and the neutral
wino behaves as the LSP with mass of several TeV, which is favored
in the thermal DM scenario \cite{Hisano:2006nn}.  
Recently, the sfermion flavor
structure is focused attention on in order to survey these models by
using indirect searches
\cite{Altmannshofer:2013lfa,Nagata:2013sba,Tanimoto:2015ota}.
In the grand unified theories (GUTs) based on the high-scale SUSY
scenarios, the specific mass spectrum yields the several features: the
gauge couplings unify at the GUT scale with higher accuracy
\cite{Hisano:2013cqa}, and the dangerous proton decay via
color-triplet Higgs exchange is suppressed due to the heavy
sfermions \cite{Hisano:2013exa,Nagata:2013sba}.

The simplest model for the high-scale SUSY scenarios is based
on the anomaly mediation \cite{Giudice:1998xp,Randall:1998uk}.  On the
other hand, the generic models may include the gauge mediated
contribution to the SUSY-breaking terms
\cite{Nelson:2002sa,Hsieh:2006ig}. In the extensions, the pattern of
gaugino masses differs from the simplest model. The vector-like
multiplets for messengers are naturally introduced since they may
obtain masses proportional to the gravitino mass via the
Giudice-Masiero mechanism \cite{Giudice:1988yz}.  If we assume that
the vector-like multiplets are in $SU(5)$ multiplets, the gauge
coupling unification is maintained in these models. Thus, 
the extensions should be considered equally to the simplest
model.

The electric dipole moments (EDMs) are important to investigate the
additional $CP$ violation in the SUSY breaking terms.  In the SM, the
EDMs for fundamental fermions are small \cite{Shabalin:1978rs}, and
thus the EDMs have high sensitivities on the new physics.  In
high-scale SUSY breaking scenarios, the generic $CP$-violating phases
are still allowed thanks to the heavy sfermions, even if the EDMs are
generated at one-loop level. However, the future experiments for EDM
searches may have sensitivities to the high-scale SUSY models.

In the high-scale SUSY scenarios, the dominant contribution to the EDMs
in the MSSM comes from the Barr-Zee two-loop diagrams, especially the
chargino/neutralino two-loop diagrams \cite{Giudice:2005rz}. When the
higgsino and wino are around a few TeV, the current experimental upper
bound on the electron EDM has already given the constraint on the
models. The ratios of electron and nucleon EDMs are predictive so that
the measurements of the ratios would lead to determination of the mass
spectrum.

On the other hand, there may exist additional contribution from gluino
in the extended models mentioned above.  The physical complex phase of
the gluino mass may arise from a relative phase of the anomaly and
gauge mediated contribution to it.  The additional $CP$-violating
source, so-called the gluino chromoelectric dipole moment (CEDM), is
induced by the physical phase of the gluino mass and the $CP$-violating
couplings of gluino and vector-like multiplets.  The nucleon EDMs are
affected by the additional source since the gluino CEDM turns into the
$CP$-violating gluon Weinberg operator \cite{Weinberg:1989dx} below
the gaugino threshold.  In this paper, we study the effect of the
gluino CEDM contribution to nucleon EDMs, and then we show the future
prospects for the observation of EDMs (electron, neutron, and proton)
in the high-scale SUSY models.

When both left- and right-handed sfermions have flavor-violating soft
mass terms, the one-loop diagrams to the EDMs are enhanced by the
heavy fermion masses. The contributions may be sizable when the flavor
violation is ${\cal O}(1)$
\cite{Hisano:2004tf,Hisano:2008hn,Fuyuto:2013gla}. However, they
are quickly suppressed when the flavor violation is small. Then, we do
not include the contribution to the EDMs from the flavor violation in
this paper.

Current status of the EDM experiments is as follows: the bounds on
the electron, neutron, and proton EDMs are given by $|d_e| < 8.7
\times 10^{-29} [e~{\rm cm}]$ \cite{Baron:2013eja}, $|d_n| < 2.9
\times 10^{-26} [e~{\rm cm}]$ \cite{Baker:2006ts}, and $|d_p| < 7.9 \times
10^{-25} [e~{\rm cm}]$, respectively.\footnote{
 The proton EDM is deduced from the Mercury EDM \cite{Griffith:2009zz}.
}
In future experiments, there are several proposals
\cite{Hewett:2012ns,Kumar:2013qya}: for instance, some neutron EDM
measurements may achieve a sensitivity of $|d_n| \sim
10^{-28}[e~{\rm cm}]$. In the proton EDM measurement at COSY
\cite{Lehrach:2012eg} and BNL \cite{Semertzidis:2011qv}, they may
achieve a sensitivity of $|d_p| \sim 10^{-29} [e~{\rm cm}]$.  For the
electron EDM, the final purpose of the ACME experiment is to reach a
sensitivity of $3 \times 10^{-31} [e\,{\rm cm}]$.  If the electron and
nucleon EDMs are discovered, we may discriminate models beyond the SM
by taking correlation among them.\footnote{ See
  Ref.~\cite{Dekens:2014jka} and the references in it for the recent
  works.}

This paper is organized as follows: in \cref{sec:model}, we introduce
the high-scale SUSY scenarios with the gauge mediation and show that
the physical phase of the gaugino mass appears.  The complex gluino mass
gives rise to the gluino CEDM, and the $CP$-violating Weinberg operator
\cite{Weinberg:1989dx} is induced by integrating out the gluino field in the
scenarios.  The gluino CEDM and the Weinberg operator in the scenarios
are shown in \cref{sec:gluinocedm}.  In the next section, we briefly
introduce our method to estimate the observable EDMs, in
particular neutron and proton EDMs. The contributions of (C)EDMs of
quarks to the nucleon EDMs are evaluated with the QCD sum rules, while
those of the Weinberg operator are based on the naive dimensional
analysis. We give the detail of calculations in
\cref{app:RGE,app:QCDsum}.  In \cref{sec:num}, we study the effect
of gluino CEDM to nucleon EDMs.  We evaluate the
electron, proton, and neutron EDMs in the
high-scale SUSY scenarios in the last of this section.  Finally, we
summarize this paper in \cref{sec:conclusion}.

\section{Complex Gaugino Mass \label{sec:model}}

In the SUSY breaking sector, we assume that there is no singlet
superfield. Under this assumption, the soft parameters are given 
as follows: mass parameters of scalar components are induced by the
Planck-suppressed higher-dimensional operators, and gaugino masses and
scalar trilinear couplings are induced with one-loop suppression by
the anomaly mediation \cite{Giudice:1998xp,Randall:1998uk}.  In
particular, the anomaly mediated gaugino masses are given by \eqs{
  M_a^{\rm AMSB} = \frac{\beta(g_a)}{g_a} m_{3/2}, } where the
subscripts $a=1$-$3$ denote the gauge groups of the SM, $U(1)_Y,
SU(2)_L$, and $SU(3)_C$.  $\beta(g_a)$ and $m_{3/2}$ denote
the beta function for the gauge coupling $g_a$ and
the gravitino mass, respectively.

We note that there also exists the additional contribution from
the higgsino-Higgs loops to the wino and bino masses.  Below the
higgsino threshold, the additional contribution is given as
\cite{Giudice:1998xp}
\eqs{
M_1^{\tilde h H} = \frac{g_1^2}{16\pi^2} \frac35 L , ~~~~~~
M_2^{\tilde h H} = \frac{g_2^2}{16\pi^2} L.
}
The parameter $L$ denotes the loop function defined as 
\eqs{
L & = \mu_H \sin 2\beta \frac{m_A^2}{|\mu_H|^2-m_A^2} \ln\frac{|\mu_H|^2}{m_A^2},
}
where $\mu_H$ and $m_A$ denote the masses of higgsino and heavy Higgs
bosons, respectively.  $\tan \beta$ denotes the ratio of the vacuum
expectation values (VEVs) of MSSM Higgs bosons.

If there exist vector-like superfields, so-called messenger multiplets
of the gauge mediation, the soft parameters differ from the simple
high-scale SUSY breaking scenario.  In order to maintain the gauge coupling unification, we assume that messenger superfields are in
$\bold{5}+\ovl{\bold{5}}$ or $\bold{10}+\ovl{\bold{10}}$
representation.  The mass terms of the messenger superfields arise
from the Giudice-Masiero mechanism \cite{Giudice:1988yz}, if the
K\"ahler potential is given as 
\eqs{
{\cal K} = |\ovl \Phi|^2 + |\Phi|^2 + \left( c_\Phi \ovl{\Phi}\Phi + \mathrm{h.c.} \right) ,
}
where $\Phi$ and $\ovl \Phi$ denote the messenger chiral superfields.
On the other hand, the superpotential $W(\Phi,\ovl\Phi)$
may have the mass terms of the messenger superfields, 
\eqs{
W(\Phi,\ovl\Phi) = M_\Phi \ovl\Phi \Phi.
}
In those cases, the mass matrix of the scalar components of $\Phi$ and
$\ovl\Phi$ is given as
\eqs{
\mathbf{m}_\phi^2 = \left(
\begin{array}{cc}
|M_\Phi + c_\Phi m_{3/2}|^2 & c_\Phi^\ast m_{3/2}^2 \\
c_\Phi m_{3/2}^2 & |M_\Phi + c_\Phi m_{3/2}|^2
\end{array}
\right)
\equiv \left(
\begin{array}{cc}
|M|^2 & -|F|e^{-i\theta_F} \\
-|F|e^{i\theta_F} & |M|^2
\end{array}
\right),
}
where the term proportional to $m_{3/2}$ arises from the Giudice-Masiero
mechanism.  For simplicity, the mass parameters are re-parametrized by
$M$ and $F$ in the last form, and $\theta_F$ denotes the complex phase
of $F$.

\begin{figure}[t]
\begin{center}
\includegraphics[width=5cm,clip]{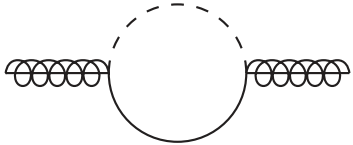}
\caption{One-loop diagram contributing to gluino mass.  Solid and
  dashed lines correspond to the propagators of fermionic
  and scalar components of chiral multiplets $\Phi, \ovl\Phi$,
  respectively.}
\label{fig:gluino_mass}
\end{center}
\end{figure}

In the above models, the gaugino masses are induced by the anomaly 
and  gauge mediation mechanisms at one-loop order.  
The anomaly mediated gluino mass is given by
\cite{Randall:1998uk,Giudice:1998xp}:
\begin{equation}
M_3^{\mathrm{AMSB}} = \frac{g_3^2}{16\pi^2}b_3 m_{3/2}\;.
\end{equation}
$b_3 = -3 + N_{\bold 5} + 3 N_{\bold{10}}$ is the coefficient for the one-loop
beta function of the $SU(3)_C$ gauge coupling,
where $N_{\bold 5}$ and $N_{\bold{10}}$ denote the numbers of pairs of
$\bold{5}+\ovl{\bold{5}}$ and $\bold{10}+\ovl{\bold{10}}$
representations, respectively.  We choose $M_3^{\mathrm{AMSB}}$ as real
for simplicity.  $M_3^{\mathrm{GMSB}}$ is induced by a diagram in \cref{fig:gluino_mass} \cite{Martin:1996zb}:
\eqs{
M_3^{\mathrm{GMSB}} &= \frac{g_3^2}{16\pi^2}(\cos\theta_F -i\sin\theta_F\gamma_5)n_3(\Phi)\left|\frac{F}{M}\right| g(x) \,, \label{eq:gluino_GMSB}
}
where $x \equiv |F/M^2|$ and $n_3(\Phi)$ is the sum of Dynkin indices
of the pair of chiral multiplets, $\Phi$ and $\ovl\Phi$.  The loop
function $g(x)$ is given by
\begin{equation}
  g(x) = \frac{(1+x)\ln(1+x) + (1-x)\ln(1-x)}{x^2}\;.
\end{equation}

For later use, we define the real gluino mass parameter
$M_{\widetilde g}$ and the phase of the gluino mass $\theta$ as:
\eqs{
M_{\widetilde g} ~ e^{i\gamma_5\theta} = M_3^{\mathrm{AMSB}} + M_3^{\mathrm{GMSB}}.
}
By the chiral rotation of the gluino field $\widetilde g^a \to
\widetilde g'^a = e^{-i\theta \gamma_5 /2} \widetilde g^a$, this
additional complex phase appears in the interaction terms between
gluino and messengers.  In next section, we show that the additional
complex phase gives rise to the gluino CEDM, and then the
$CP$-violating Weinberg operator also arises from the gluino
CEDM operator below the gluino threshold scale.

\section{Gluino CEDM and Weinberg Operator \label{sec:gluinocedm}}

\begin{figure}[t]
\begin{center}
\includegraphics[width=5cm,clip]{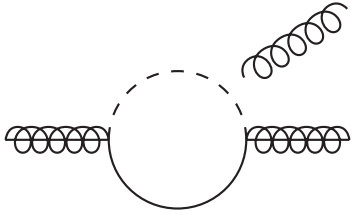}
\caption{One-loop diagram inducing gluino CEDM. 
Solid and dashed lines correspond to propagators of fermionic and scalar components of chiral multiplets $\Phi, \ovl\Phi$, respectively.}
\label{fig:gluino_CEDM}
\end{center}
\end{figure}

In our models,  the gluino CEDM is generated by the messenger loop
diagrams (\cref{fig:gluino_CEDM}) since there exists the non-vanishing
$CP$-violating phase in the gluino-messenger interaction after the chiral rotation of the gluino.  
The gluino CEDM $\widetilde d_{\widetilde g} $ is given as 
\eqs{ {\cal L}_{\widetilde g~\text{CEDM}} = - \frac{i}{4}
  \widetilde d_{\widetilde g} \ovl{\tilde g^b}\si^{\mu\nu}\gamma_5
  G^a_{\mu\nu} [T^a]_{bc} \tilde g^c\;, }
where $\si^{\mu\nu} =
\frac{i}{2}[\gamma^\mu,\gamma^\nu]$ and $G^a_{\mu\nu} = \partial_\mu
G^a_\nu - \partial_\nu G^a_\mu + g_3f^{abc}G^b_\mu G^c_\nu$.
$[T^a]_{bc} = - i f^{abc}$ and $f^{abc}$ is the structure constant for
the $SU(3)_C$.  

We estimate the relevant $CP$-violating terms at the
gluino mass scale ($M_{\widetilde g}$) from those at the messenger
mass scale ($M_{\rm mess}$) by using the renormalization group
equation (RGE) analysis.  It is useful to define the dimension-six
gluino CEDM operator in order to estimate the RGE evolution.  The
gluino CEDM operator $\mathcal{O}_{\widetilde g}$ and its Wilson
coefficient $C_{\widetilde g}$ are defined as
\begin{equation}
\mathcal{O}_{\widetilde g} = \frac{1}{4}M_{\widetilde g}g_3f^{abc} \ovl{\tilde g^a}\si^{\mu\nu}\gamma_5\tilde g^cG^b_{\mu\nu}\;, ~~~
\widetilde d_{\widetilde g}  = M_{\widetilde g} g_3 C_{\widetilde g}.
\end{equation}
By evaluating a diagram in \cref{fig:gluino_CEDM}, we obtain the Wilson coefficient of $\mathcal{O}_{\widetilde g}$ as
\begin{equation}
C_{\widetilde g}(M_{\mathrm{mess}}) = - \frac{g_3^2}{32\pi^2}\frac{1}{M_{\widetilde g}}\frac{M}{m_+^2}\sin(\theta + \theta_F)\left[A(r_+) + B(r_+)\right] - (m_+,r_+ \rightarrow m_-, r_-),
\end{equation}
where $m_\pm^2 = |M|^2 \pm |F|$ are the mass eigenvalues of the mass
matrix for the scalar components of $\Phi$ and $\ovl\Phi$, and $r_\pm
= |M|^2/m_\pm^2$.  $\theta$ and $\theta_F$ are respectively the phases
of the complex gluino mass and the off-diagonal element $F$ of the
mass matrix $\mathbf{m}_\phi^2$, as defined in the previous section.
The loop functions $A(r)$ and $B(r)$ are given as
\eqs{
A(r) \equiv \frac{1}{2(1-r)^2}\left(3-r+\frac{2\ln r}{1-r}\right)\;, ~~
B(r) \equiv \frac{1}{2(1-r)^2}\left(1+r+\frac{2r\ln r}{1-r}\right)\;.
}

Now, we estimate the gluino CEDM at the gluino mass scale by using the
RGEs between the messenger and the gluino mass scales.  The RGE for
the Wilson coefficient $C_{\widetilde g}(\mu)$ at the leading order is
given as
\begin{equation}
\pdiff{1}{}{\ln\mu}C_{\widetilde g}(\mu) = \frac{g_3^2}{16\pi^2}\gamma_{\mathcal{O}_{\widetilde g}}C_{\widetilde g}(\mu)\;,
\end{equation}
where $\gamma_{\mathcal{O}_{\widetilde g}} = 12N_C$ and $N_C(=3)$ is
the number of colors.  This anomalous dimension is found by
substituting the Casimir invariant $C_F = N_C$ for $C_F = 4/3$ in the
anomalous dimension of the dipole operator for $b \to s g$
\cite{Ciuchini:1993fk,Degrassi:2005zd}.  The gluino CEDM at the gluino
mass scale is obtained as follows:
\eqs{
\frac{\widetilde d_{\widetilde g}(M_{\mathrm{mess}})}{\widetilde d_{\widetilde g}(M_{\widetilde g})} = \left(\frac{\alpha_s(M_{\mathrm{mess}})}{\alpha_s(M_{\widetilde g})}\right)^{\gamma_{\mathcal{O}_{\widetilde g}}/2b_3-3N_C/b_3+1/2}\;.
}
Here, $b_3 = -7|_{\mathrm{SM}} + 2|_{\mathrm{gluino}}$ where the
subscripts ``SM" and ``gluino" indicate the contribution from the SM
particles and gluino, respectively.
The exponents $-3N_C/b_3$ and $1/2$ are due to the one-loop renormalization-group evolution of the gluino mass $M_{\widetilde g}$ and the strong gauge coupling $g_3$, respectively.

 \begin{figure}[t]
\begin{center}
\includegraphics[width=4cm,clip]{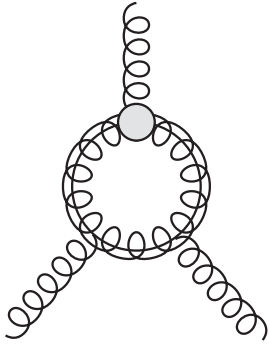}
\caption{One-loop diagram inducing $CP$-violating Weinberg operator. The blob denotes the gluino CEDM operator.}
\label{fig:Weinberg_op}
\end{center}
\end{figure}
 
The $CP$-violating Weinberg operator is induced through the gluino
one-loop diagram in \cref{fig:Weinberg_op}.  The
effective Lagrangian for the Weinberg operator is defined
as \cite{Weinberg:1989dx}
\eqs{
{\cal L}_W = C_W {\cal O}_W, ~~~~~
\mathcal{O}_W = -\frac{1}{6} g_3 f^{abc} \ep^{\mu\nu\rho\si}G^a_{\mu\lam}G^{b}_\nu{}^\lam G^{c}_{\rho\si}\;.
\label{eq:weinberg_ceff}
}
Here, $\ep^{\mu\nu\rho\si}$ is the totally-antisymmetric tensor with
$\ep^{0123}=+1$.  At the gluino mass scale, we obtain the Wilson
coefficient of the Weinberg operator by matching the amplitude as
follows \footnote{
In Refs~ \cite{Kamenik:2011dk,Brod:2013cka,Sala:2013osa,Gorbahn:2014sha}, authors considered the similar effects arising from the $CP$-violating coupling of heavy quarks.
}:
\begin{equation}
C_W(M_{\widetilde g}) = \frac{N_C g_3^2}{32\pi^2}C_{\widetilde g}(M_{\widetilde g})\;. \label{eq:CW_match}
\end{equation}

Notice that quark (C)EDMs are induced as a three-loop contribution after integrating out the messenger multiplets.
The Weinberg operator could be also induced directly after
squarks and messenger multiplets are integrated out, not via the
gluino CEDM. These contributions are suppressed by the masses of
squarks and messenger multiplets. Thus, the gluino CEDM contribution
is dominant for the Weinberg operator as far as gluino is lighter than
them.

\section{Nucleon Electric Dipole Moments\label{sec:nucleonedm}}

In previous section, we show the expression of the $CP$-violating
Weinberg operator induced by the gluino CEDM at the gluino mass scale.  The
quark EDMs are also induced through the Barr-Zee diagrams
\cite{Barr:1990vd}, which are dominated by only the chargino and
neutralino loops in the MSSM based on the high-scale SUSY scenarios
\cite{Giudice:2005rz}.  The quark CEDMs are radiatively induced from
the Weinberg operator so that they are subdominant.

In this section, we summarize the RGE evolutions of the $CP$-violating
operators and results of the QCD sum rules and the naive dimensional
analysis in order to obtain the nucleon EDMs at the hadron scale ($\mu
= 1\;\mathrm{~GeV}$) in a compressed way.

The $CP$-violating operators in the QCD sector below the gluino mass scale
are given as
\eqs{
\mathcal{L}_{\cancel{\mathrm{CP}}} &= \ovl \theta\frac{g_3^2}{32\pi^2}G^a_{\mu\nu}\widetilde{G}^{a,\mu\nu} \\
&\quad -\frac{i}{2}\sum_{q=u,d,s}d_q\ovl{q}(F\cdot\si)\gamma_5q 
- \frac{i}{2}\sum_{q=u,d,s}\widetilde {d}_q g_3\ovl q (G\cdot\si)\gamma_5q \\
&\quad +\frac{1}{3}wf^{abc}G_{\mu\nu}^a\widetilde{G}^{b,\nu\rho}G^{c}{}_\rho{}^\mu\;. \label{eq:CP_viol}
}
Here, $F_{\mu\nu}$ and $G^a_{\mu\nu}$ are the electromagnetic and
gluon field strength tensors, respectively, and we define as $F\cdot
\si = F_{\mu\nu}\si^{\mu\nu}$ and $G\cdot\si =
G^a_{\mu\nu}\si^{\mu\nu}T^a$.  The dual field strength tensor is defined as
$\widetilde{G}^{a,\mu\nu} = \frac12 \ep^{\mu\nu\rho\si} G^a_{\rho\si}$.
The first term of \cref{eq:CP_viol} is the dimension-four $CP$-violating term, 
so-called QCD $\theta$-term.
Since this operator, however, does not mix with the other operators,
we neglect the QCD $\theta$-term in the RGE analysis.
The second and third terms of \cref{eq:CP_viol} correspond to the quark EDMs and CEDMs, respectively. 
The last term is the Weinberg operator \cite{Weinberg:1989dx}.
The coefficient of the Weinberg operator $w$ is given as $w = -g_3 C_W$, which 
is evaluated in the previous section.

In our numerical evaluation of the nucleon EDMs, we include the RGE
evolutions of these operators between the gluino mass scale and the scale
of $\mu = 1~{\rm GeV}$.  The RGEs for the Wilson coefficients at the
leading order are given by Ref.~\cite{Degrassi:2005zd}.  The detail of
RGEs with mixing of the quark (C)EDMs and the $CP$-violating Weinberg
operator is given in \cref{app:RGE}.

Next, we show the nucleon EDMs induced by the quark (C)EDMs via the
QCD sum rules.  In Ref.~\cite{Hisano:2012sc}, the neutron EDM $d_n$ is
related to the quark (C)EDMs by using the QCD sum rules
at the renormalization scale $\mu = 1~{\rm GeV}$.  Similarly, the
proton EDM $d_p$ is also associated with the quark (C)EDMs at the
scale of $\mu = 1~{\rm GeV}$.  We obtain the relation between the
EDMs for light nucleons and the quark (C)EDMs as follows:
\eqs{
d_p & = - 1.2 \times 10^{-16}[e~{\rm cm}] \ovl\theta 
+ 0.78 d_u- 0.20d_d
+ e( -0.28 \widetilde d_u + 0.28 \widetilde d_d + 0.021 \widetilde d_s), \\
d_n & = 8.2 \times 10^{-17}[e~{\rm cm}] \ovl\theta 
- 0.12 d_u + 0.78 d_d
+ e( -0.30 \widetilde d_u + 0.30 \widetilde d_d - 0.014 \widetilde d_s). \label{eq:EDMs_1GeV} \\
}
The explicit formulae and the numerical values are presented in
\cref{app:QCDsum}.  The quark EDM contributions to the neutron EDM are
consistent with the recent result with the lattice QCD simulation 
\cite{Bhattacharya:2015esa}. 
In the following numerical analyses, we use the results from the QCD
sum rules with $\ovl\theta = 0$\footnote{ If we impose the Peccei-Quinn
  symmetry, the theta parameter $\ovl \theta$ is induced, and the
  formulae for the nucleon EDMs are changed (the explicit expressions
  are given in the \cref{app:QCDsum}), especially the coefficients of
  $\widetilde d_q$.  However, in our study, the quark CEDMs are
  subdominant since they are induced only through the RGEs, and
  our results are almost unchanged in each case.}.

The nucleon EDMs induced by the Weinberg operator are given by \cite{Demir:2002gg}:
\begin{equation}
d_N(w) \sim e(10-30)\;\mathrm{MeV} \;w (1\,{\rm GeV}),~~~(N=n,p). \label{eq:w_1GeV}
\end{equation}
This is based on the naive dimensional analysis. The sign of the
contribution of the Weinberg operator is also ambiguous. We adopt
the value $d_N(w) / e= 20\; \mathrm{MeV} \;w (1\,{\rm GeV})$ as the
nucleon EDMs induced by Weinberg operator in the following numerical
analyses.

\section{Numerical Results\label{sec:num}}

\begin{figure}[t]
\centering
\includegraphics[width=5cm,clip]{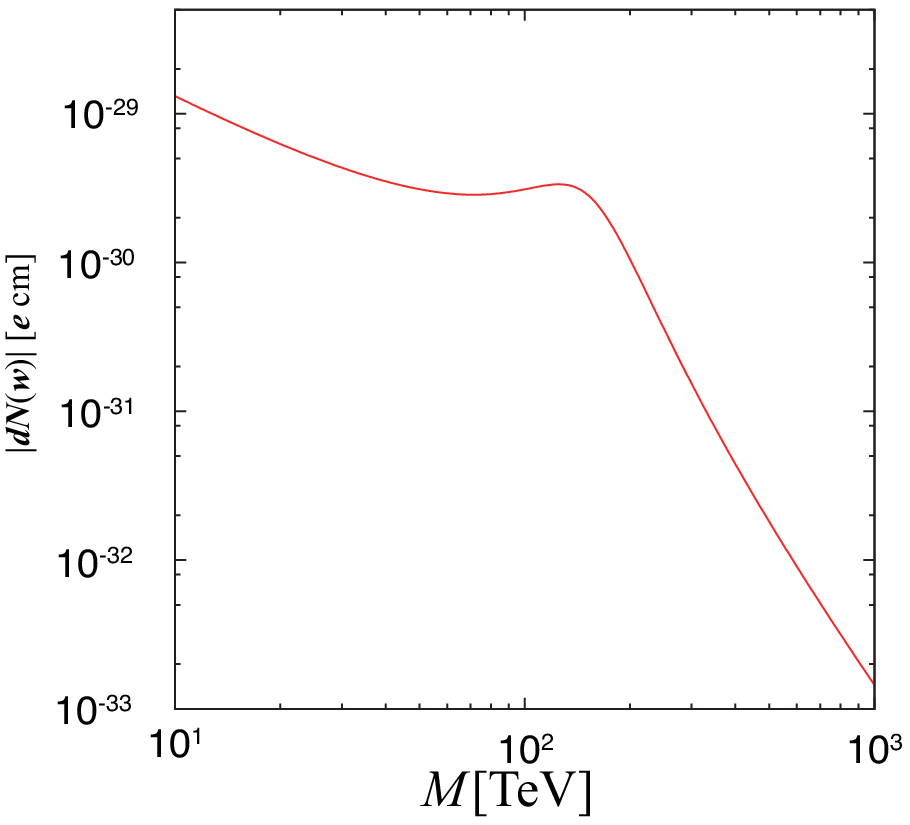}
\includegraphics[width=5cm,clip]{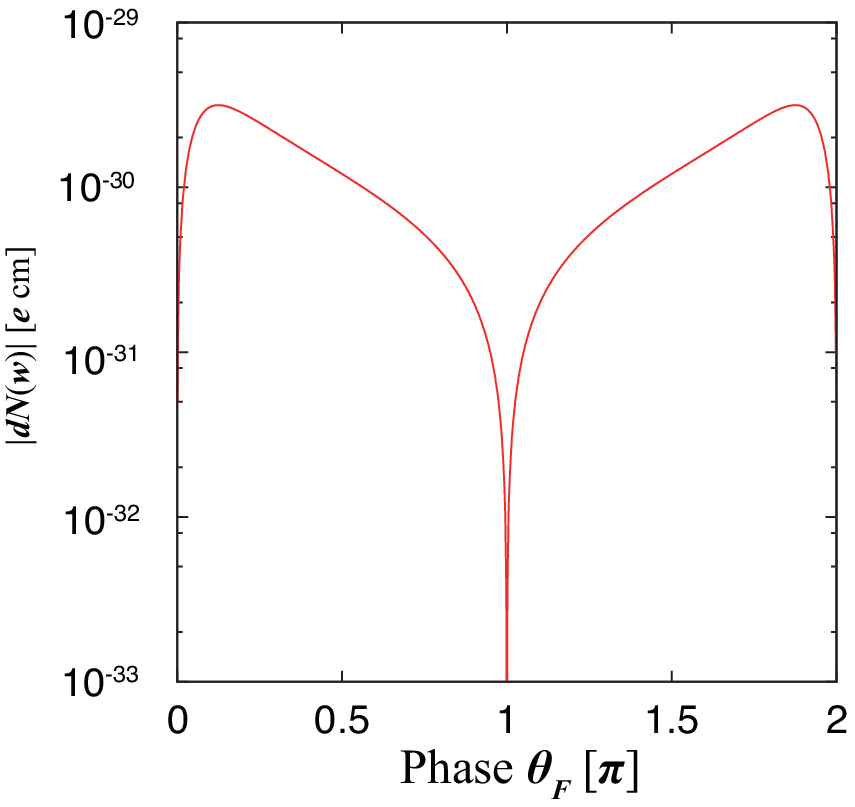}
\includegraphics[width=5cm,clip]{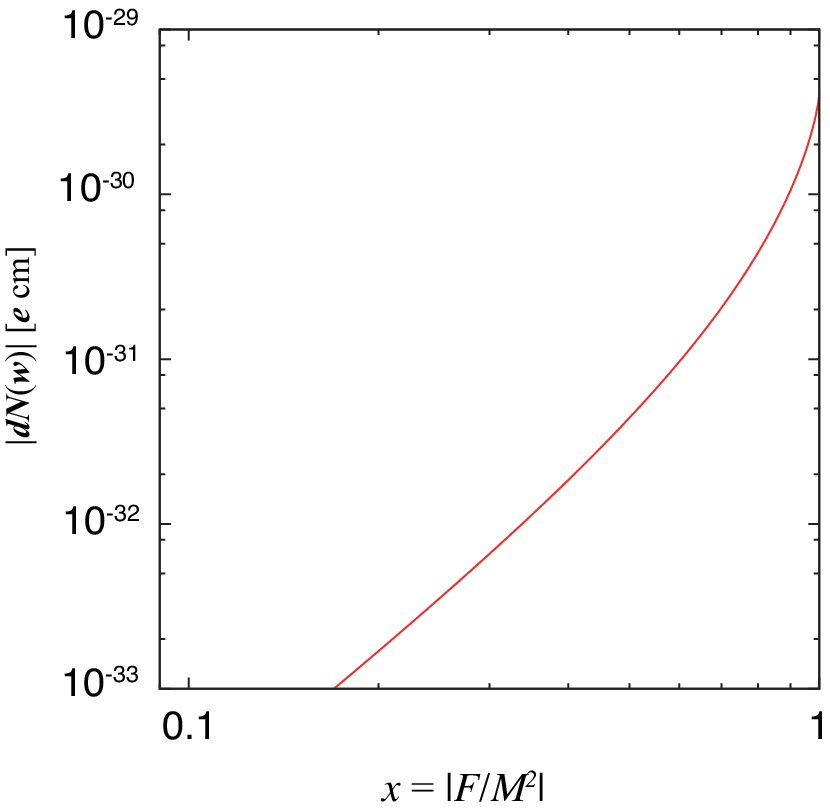}
\caption{Parameter dependence of $d_N(w)$ with fixed sfermion mass scale ($M_S=100\,{\rm TeV}$). 
(Left): Messenger mass $M$ dependence with $\theta_F = 0.125\pi$ and $x=0.99$.
(Middle): Phase $\theta_F$ dependence with $M=M_S$ and $x=0.99$.
(Right): $x \, (\equiv |F/M^2|)$ dependence with $\theta_F = 0.125\pi$ and $M=M_S$.
}
\label{fig:weinberg_properties}
\end{figure}

Now, we estimate the electron and nucleon EDMs in the high-scale SUSY
scenarios.

To begin with, let us consider the parameter dependence of nucleon EDMs
induced by the gluino CEDM.  In this evaluation, we assume that
sfermions, heavy Higgs bosons and the gravitino are degenerate in
 mass $M_S$, and we take $M_S = 100~{\rm TeV}$.  Once we fix $M_S$, we have three
parameters; $M$ and $|F|$ which are the mass parameters of the scalar fields
of messengers, and $\theta_F$ which is the complex phase of $F$.  In
the following numerical analyses, we choose $M$,
$\theta_F$, and $x \, (\equiv |F/M^2|)$ as independent parameters.

\cref{fig:weinberg_properties} shows parameter dependence of the
nucleon EDMs $d_N$ $(N=p,n)$ induced by the gluino CEDM.  In the middle panel of
\cref{fig:weinberg_properties}, we show the $\theta_F$ dependence of
gluino-induced nucleon EDMs. In this figure, we set $M = 100~{\rm
  TeV}$ and $x = 0.99$. If $\theta_F = 0$ or $\pi$, there is no
contribution from the gluino-induced nucleon EDM, since $M^{\rm GMSB}$
and the couplings of the gluino-messenger interaction are real.  The
maximal contribution is given by $\theta_F \sim 0.125 \pi$.  When the
real and imaginary parts of the gluino mass are comparable to each other,
the physical phase of the gluino mass $\theta$ is maximized.  Since the
coefficient of the Weinberg operator is proportional to $\sin
(\theta+\theta_F)/M_{\widetilde g}$, the maximum contribution arises
when $\theta+\theta_F \sim \pi/2$ and $M_{\widetilde g}$ is small.

In the right of \cref{fig:weinberg_properties}, the $x$ dependence of
nucleon EDM is shown.  This dependence is evaluated with $M = 100~{\rm
  TeV}$ and $\theta_F = 0.125 \pi$. The gluino CEDM approaches to the
maximum as $x \to 1$.  On the other hand, one of the scalars of
messengers becomes massless when $x=1$.  In order to avoid this
situation, in the following calculations, we set $x=0.99$ for
simplicity.

Finally, we show the $M$ dependence of nucleon EDM in
the left panel of \cref{fig:weinberg_properties}.  In this figure, we
set $x = 0.99$ and $\theta_F = 0.125 \pi$.  The Wilson coefficient of
Weinberg operator (\cref{eq:CW_match}) behaves as follows: \eqs{ C_W
  \propto
\begin{cases}
\dps \frac{1}{M_S M} \sin(\theta+\theta_F) & (M \ll M_S),
 \\ 
\dps \frac{1}{M^2} \sin(\theta+\theta_F) & (M_S \ll M). \label{eq:cw_behavior}\\
\end{cases}
} 
When $M\gg M_S$, the gluino mass mainly comes from the gauge mediation
so that the Weinberg operator is highly suppressed by $M^2$ and also
the suppressed $CP$ phase in $\sin(\theta+\theta_F)$.  If $M\ll M_S$,
the gluino mass is dominated by the anomaly mediated contribution, and
thus, the Weinberg operator is suppressed by $M_S M$.  On the other
hand, in the region $M \sim M_S$, the nucleon EDMs are slightly enhanced
since the gauge mediated gluino mass is comparable to $M^{\rm AMSB}$
and thus the $CP$ phase of the gluino mass is maximal.

Now, let us compare the nucleon and electron EDMs in the high-scale
SUSY scenarios, in which the Barr-Zee diagrams and gluino CEDM
contribute to the EDMs. In the high-scale SUSY scenarios, all MSSM
scalar particles except the SM Higgs multiplet are assumed to be
heavy. Within the MSSM, the main contributions to the electron and
nucleon EDMs arise from the Barr-Zee type two-loop diagram contributions
\cite{Giudice:2005rz}. The diagrams include loops of charginos and
neutralinos, and the contributions to EDMs are suppressed by $m_f/M_2
\mu_H$\footnote{
This behavior of the Barr-Zee contribution is understood by using the 
effective theory of wino after integrating out higgsino \cite{Hisano:2014kua}.}.
 (The electron or quark mass $m_f$ appears in the quark and electron EDMs 
due to the chiral nature.)
Since the higgsino mass $\mu_H$ is a model-dependent parameter, we
perform the evaluation of EDMs in a range of $\mu_H/M_S \in
[10^{-2},1]$.  The $CP$ phase of $\mu_H$ is also model-dependent and
therefore we set the $CP$ phase in the mass matrices of charginos and
neutralinos to make the Barr-Zee contributions maximized \footnote{ In
  the Ref.~\cite{Giudice:2005rz}, there exist two independent phases
  $\phi_1$ and $\phi_2$, which are the combinations of phases of
  gaugino masses, gaugino-higgsino-Higgs couplings, and $\mu_H$.  We
  set $\sin \phi_1 = \sin \phi_2 = 1$ in our numerical analyses,
  unless otherwise stated. 
However, since there remain the two ambiguities to determine the nucleon EDMs.
One is the sign ambiguity in the contribution of the Weinberg operator to the nucleon EDMs.
The other is the choice of the independent phases as $\sin \phi_1 = \sin \phi_2 = -1$, 
which is also allowed to maximize the absolute value of the Barr-Zee contribution.
There is no need to care about these ambiguities as far as discussing absolute values for each contributions.  
}.  We note that we set $\tan \beta
= 3$ since $\tan \beta$ is also model-dependent parameter \footnote{We
  do not take care of whether the observed Higgs mass is realized in
  the parameter set in this paper, though we set the Higgs mass in the
  Barr-Zee contributions to be the observed value.  If there exist
  additional matters contributing to the Higgs mass (such as
  Ref.~\cite{Evans:2014xpa}), the favored values for $\tan\beta$ to
  realize the observed Higgs mass may differ from those in the simple
  high-scale SUSY models \cite{Giudice:2011cg,Ibe:2011aa}.  }.

In the following numerical analyses, we assume that all MSSM scalar
particles except the SM Higgs boson have the same mass $M_S$ and three
parameters which are associated with the gluino induced nucleon EDMs
are set to be $M = M_S$, $\theta_F = 0.125 \pi$, and $x = 0.99$ in order
to study the maximal gluino CEDM effects.  The higgsino mass $\mu_H$ is
estimated as follows: input value for the higgsino mass is given at
the renormalization scale $\mu = M_S$, and then, we estimate the
higgsino mass at $\mu = \mu_H(M_S)$ by using the one-loop RGEs for
higgsino and gauginos.

\begin{figure}[t]
\centering
\includegraphics[width=7cm,clip]{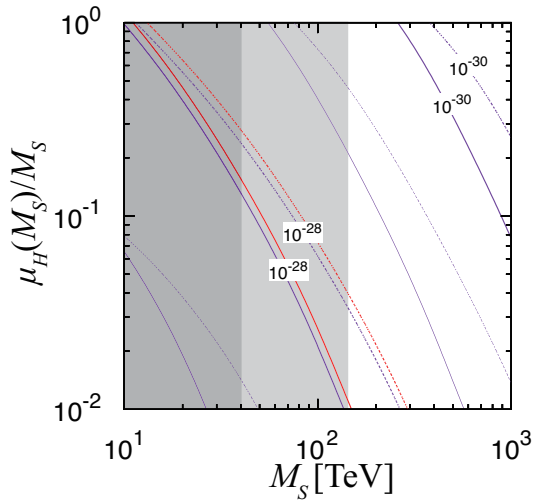}
\caption{Electron EDM in high-scale SUSY models with $N_{\bf 5} = 0$ (dotted) and $1$ (solid). 
Red lines correspond to the current bound; $|d_e| < 8.7 \times 10^{-29} ~ [e\,{\rm cm}]$ \cite{Baron:2013eja}. 
Dark gray and light gray regions are excluded by gluino search at the LHC with $N_{\bf 5} = 0$ and $1$, respectively.} 
\label{fig:de}
\end{figure}

The Barr-Zee contributions are estimated as follows: the input
parameters for the chargino and neutralino mass matrices are estimated
at $\mu = \mu_H(M_S)$, the coupling constants associated with the
chargino and neutralino loops are also estimated at $\mu
=\mu_H(M_S)$, and other couplings are estimated at the EW
scale.

First, we show electron EDM $d_e$ in the high-scale SUSY scenarios (\cref{fig:de}).  
We assume that there is no messenger superfield ($N_{\bf 5}=0$; dotted
lines in \cref{fig:de}) and that there is messenger superfields
($N_{\bf 5}=1$; solid lines in \cref{fig:de}).  The dark (light)
shaded regions are excluded by the gluino search at the LHC
\cite{Aad:2014wea,Chatrchyan:2014lfa}, that is, $M_{\widetilde g}^{\rm
  pole} < 1.3\ {\rm TeV}$.  The red
lines correspond to the current bound on the electron EDM measured by
the ACME experiment, $|d_e| < 8.7 \times 10^{-28} e\,{\rm cm}$.  It is
found that the future experiment for the electron EDM may have
sensitivities to the SUSY breaking scale of $M_S \sim 10^3~{\rm TeV}$ in
each scenario.

In the high-scale SUSY models with messengers, there are several
differences from the models with no messengers.  One is that the constraint
on the SUSY breaking scale becomes severe in the models with
messengers.  This is because that the cancellation between
the anomaly and gauge mediated contributions reduce the gluino mass
since we choose parameters to maximize the gluino CEDM.  Another is that the
Barr-Zee contributions are slightly suppressed since the masses
of wino and bino become large in the extended models.  

\begin{figure}[t]
\centering
\includegraphics[width=6.5cm,clip]{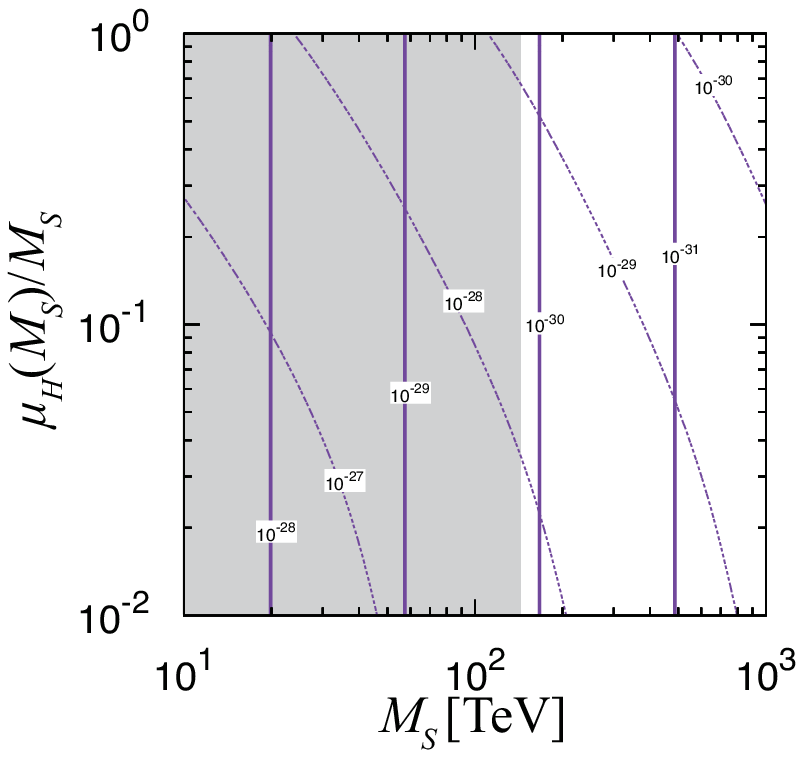}
\includegraphics[width=6.5cm,clip]{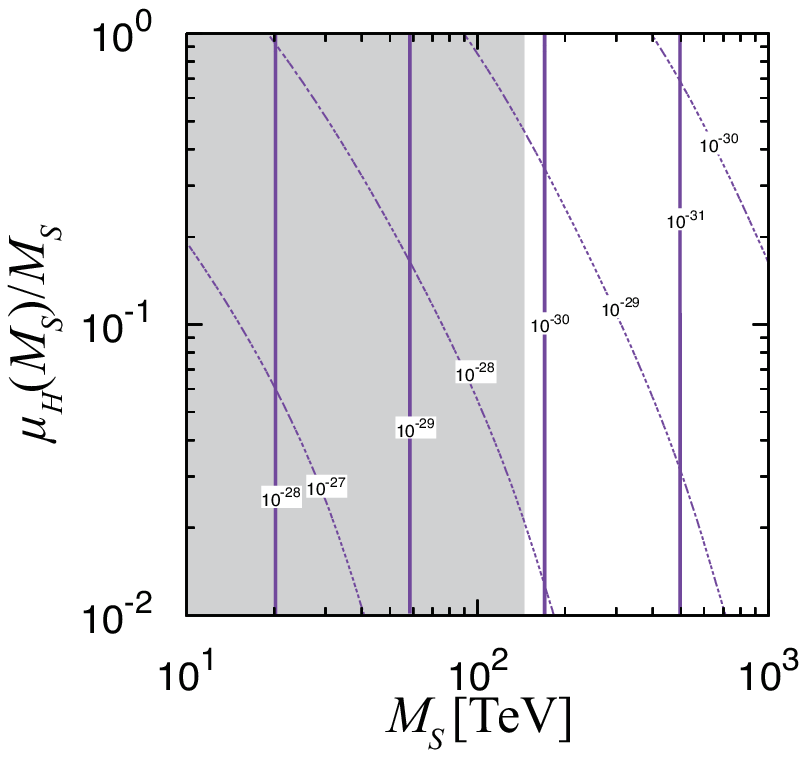}
\caption{
Nucleon EDMs from Barr-Zee diagrams and gluino CEDM.
Neutron (proton) EDM is shown in left (right) panel.
Dotted lines describe the case of vanishing gluino CEDM, that is, we set $\theta_F = 0$.
Solid lines correspond to the case of vanishing Barr-Zee contribution.
Again, gray regions in each figures are excluded by gluino search at the LHC.
In each panels, number of messengers is set to be $N_{\bf 5} = 1$.
}
\label{fig:dnnoBZ}
\end{figure}

Next we investigate the nucleon EDMs
in the high-scale SUSY models.  First, we compare the nucleon EDMs
induced by only the gluino CEDM with the Barr-Zee 
contribution.  In \cref{fig:dnnoBZ}, we show the nucleon EDMs induced by only
the gluino CEDM (solid lines) or the Barr-Zee
contribution (dotted lines).  The neutron (proton)
EDM is shown in the left (right) panel of \cref{fig:dnnoBZ}.  If the
Barr-Zee contributions vanish, the nucleon EDMs are
induced by only the gluino CEDM.  The nucleon EDMs induced by
only the gluino CEDM are almost the same even if the quark (C)EDMs are
induced via the RG mixing.
On the other hand, if there is no physical phase of the gluino mass,
the nucleon EDMs come from the Barr-Zee contributions.  The
Barr-Zee contributions are suppressed in
the region of heavy higgsino and gauginos.  Since there remains the 
ambiguity in the relative sign between the Barr-Zee contribution and the Weinberg
operator, the nucleon EDMs are not determined if there exist both
contributions.  However, it is found that these contributions may be
comparable in some region, especially heavy higgsino.  Thus, we need
to include the gluino CEDM contribution to determine precisely the
nucleon EDMs.

\begin{figure}[t]
\centering
\includegraphics[width=6cm,clip]{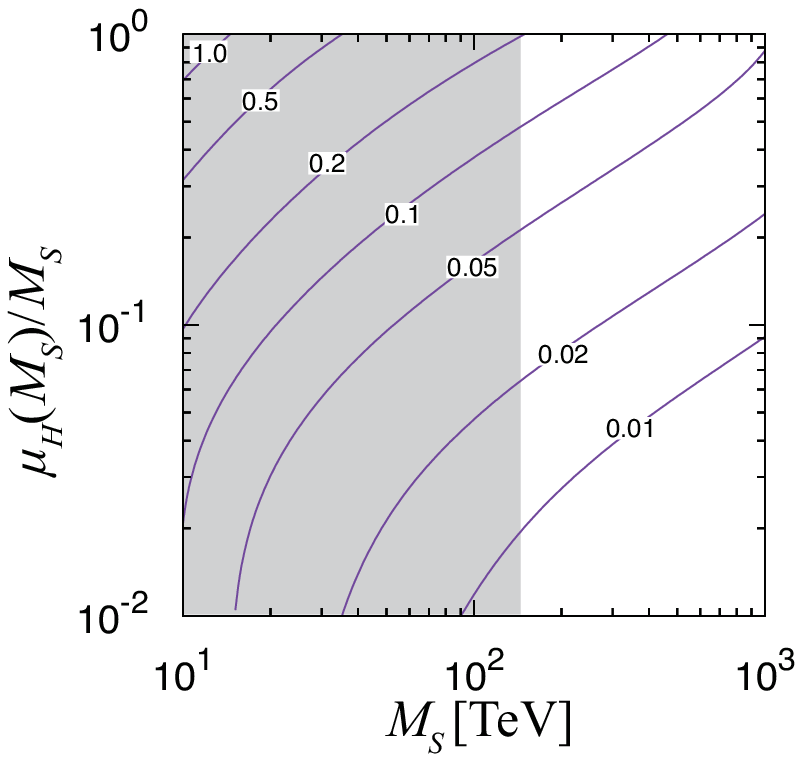}
\includegraphics[width=6cm,clip]{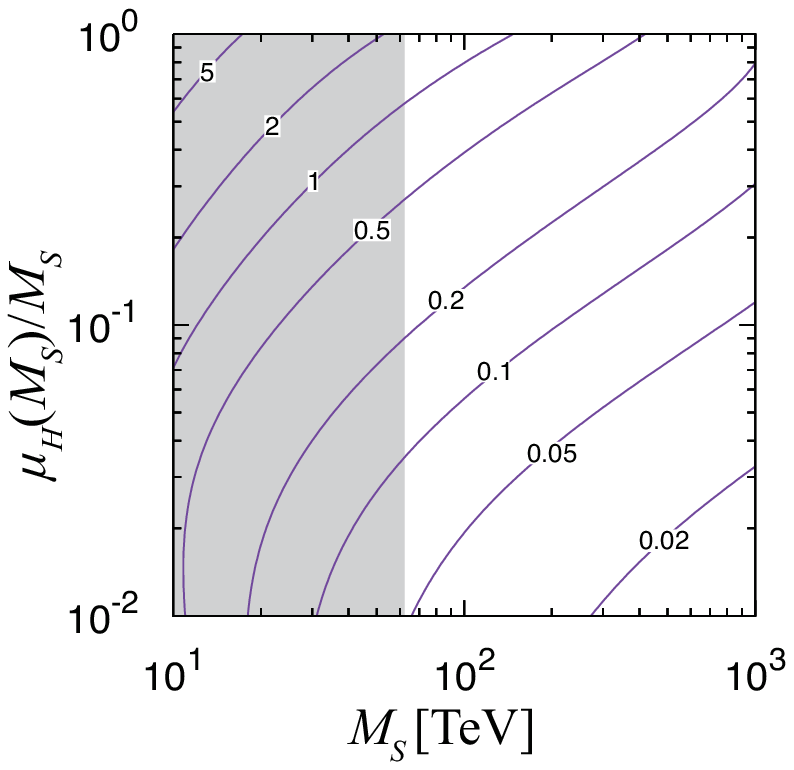}
\caption{
Ratio of $d_{n{\rm W}}/d_{n{\rm BZ}}$ with mass of messengers to be $M_S$ ($0.1M_S$) in left (right) figure.
Shades in each figures show excluded region by gluino search at the LHC, again.
In each figures, we set number of messengers to be $N_{\bf 5} = 1$.
}
\label{fig:ratio_dwdN}
\end{figure}

Now, let us estimate the ratio of neutron EDMs induced by the Barr-Zee
diagrams and the Weinberg operator.  In this evaluation, we fix the
phases of Barr-Zee contributions as mentioned above, and we also 
estimate the quark (C)EDMs and the Weinberg operator at 1\,GeV with
the RGEs.  In \cref{fig:ratio_dwdN}, we show
the ratios of the neutron EDMs induced by the Weinberg operator ($d_{n{\rm W}}$) 
and induced by the Barr-Zee diagrams ($d_{n{\rm BZ}}$).
Since the quark (C)EDMs and the Weinberg operator mix with each other,
it is not obvious to discriminate between the neutron EDM derived from quark (C)EDMs and that derived from the Weinberg operator at low energy.
We, however, identify the neutron EDM induced by the Barr-Zee 
contribution ($d_{n{\rm BZ}}$) with $d_n$ defined in
\cref{eq:EDMs_1GeV}, which is dominated by the Barr-Zee contribution
since the RGE effect is negligible due to the one-loop
suppression.  Similarly, one induced by the Weinberg operator
($d_{n{\rm W}}$) is identified with $d_n(w)$ defined in
\cref{eq:w_1GeV}.

In the left figure of \cref{fig:ratio_dwdN}, we show the ratio
$|d_{n{\rm W}}/d_{n{\rm BZ}}|$, taking the messenger mass to be $M_S$.
(We will discuss the right figure of \cref{fig:ratio_dwdN} below.)
In the large $\mu_H$ limit, the Barr-Zee contributions are suppressed
since the EDMs induced by the Barr-Zee diagrams proportional to $d_f
\sim m_f/M_2 \mu_H$ as mentioned above.  On the other hand, in the
small $M_S$ region, the neutron EDM induced by the gluino CEDM is
enhanced due to the smallness of the gluino mass. Thus, in the region
of small $M_S$ and large higgsino mass, the dominant contribution
arises from gluino CEDM, though such a region is constrained by the
gluino search at the LHC.

In the last of this section, let us consider the case of light messengers.  
The light messengers would realized if some symmetry is imposed to the messengers. 
As mentioned in the beginning of this section, the gluino CEDM is induced by the $CP$-violating coupling of gluino and messengers since the gluino mass is dominated by the anomaly mediated one.  
If the messengers have the mass lighter than $M_S$, the suppression of the gluino CEDM becomes mild.  
Therefore, the nucleon EDMs induced by the gluino CEDM become large in comparison with the heavy messenger case.

In the light messenger case, since the gaugino masses are dominated by
the anomaly mediated contribution, the phase of the gluino mass $\theta$ in
the case is approximately zero, and then the gluino CEDM
 is proportional to $\sin \theta_F$.  Thus, in the following
numerical evaluation, we set $\theta_F = \pi/2$ in order to maximize
the gluino CEDM.  If the supersymmetric masses of messengers
are sufficiently light, the lightest scalars of
messengers may have masses much lighter than the EW scale. In the following analysis,
the scale of messengers is set to be $M = 0.1 M_S$ in order to avoid
too light scalars of messengers while 
$x$ is 0.99.

\begin{figure}[t]
\centering
\includegraphics[width=6.5cm,clip]{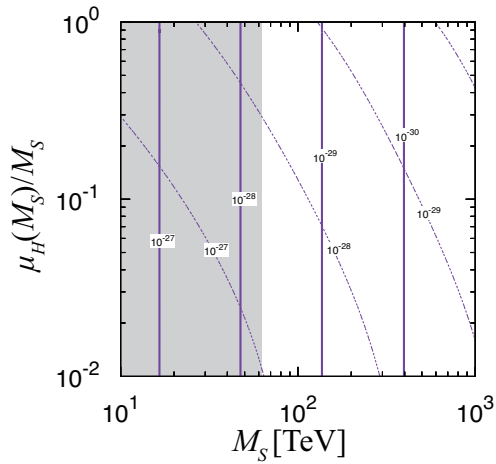}
\caption{ 
Neutron EDMs from the Barr-Zee diagrams and the gluino CEDM in the light messenger case ($M = 0.1 M_S$).  
Dotted lines describe the case of no gluino CEDM.  
Solid lines correspond to the case of no Barr-Zee contribution.  
The gray regions in each figures are excluded by the gluino search at the LHC.
The number of messengers is set to be $N_{\bf 5} = 1$.  
}
\label{fig:light_scenario_dn}
\end{figure}

We show the neutron EDM in the light messenger case in \cref{fig:light_scenario_dn}.  
The dotted and solid lines respectively describe the cases of no gluino CEDM and no Barr-Zee contribution.  
In this case, the neutron EDM via the gluino CEDM become larger than in the case of heavy messengers, as expected.  
On the other hand, the Barr-Zee contributions are the same in size as the case of heavy messengers.
Therefore, the gluino CEDM gives the sizable contribution to the neutron EDM.
If there exist both of the Barr-Zee and gluino CEDM contributions, the nucleon EDMs may be enhanced or suppressed.
However, we do not evaluate the nucleon EDMs in this case since we have the sign ambiguity for the Weinberg operator in the nucleon EDMs.

As is the case for the heavy messenger, we also show the ratio of each contribution with light messengers (in the right figure of \cref{fig:ratio_dwdN}).  
In this case ($M<M_S$), the contribution from the Weinberg operator is enhanced due to the light messenger mass as mentioned in \cref{eq:cw_behavior}.  
Thus, the $d_{n{\rm W}}$ dominates the neutron EDMs in a broad region in the light messenger scenario.

In \cref{fig:dnnoBZ}, we have shown that the proton EDMs induced by the Barr-Zee diagrams and the gluino CEDM behave similar as the neutron EDMs.
The behavior of the proton EDM is also similar to the neutron EDM in the light messenger scenario.
Since 
the future experiments of the proton EDM may have sensitivities to $|d_p| \sim 10^{-29}e~{\rm cm}$, the gluino CEDM effect via the proton EDM may be found.

Before we conclude this paper, we mention the uncertainty of numerical calculation.  
The error of $\alpha_s(M_Z) = 0.1185 \pm 0.0006$ gives uncertainty on the results about ${\cal O}(1)$\% and then it does not appear in the numerical analyses.
Thus, the uncertainties mainly come from the QCD sum rules and the naive dimensional analysis which we use in order to obtain the nucleon EDMs from the quark (C)EDMs and the Weinberg operator. 
Especially, the uncertainty in the naive dimensional analysis should be large so that we might expect larger  contributions from the gluino CEDM to the nucleon EDMs. 
Furthermore, in the above, we assume $N_{\bf 5} = 1$ and $N_{\bf 10} = 0$ for simplicity. 
If more messenger fields are introduced, larger contribution from the gluino CEDM to the nucleon EDMs is expected if the $CP$ phases are aligned to be constructive.

\section{Conclusion and Discussion \label{sec:conclusion}}

In this paper, we estimated the nucleon and electron EDMs in the
high-scale SUSY models.  Even if the gaugino masses induced by the
anomaly mediation are real, the additional contributions to gaugino
masses may give rise to the physical phases of gaugino masses.  We show the
case in the high-scale SUSY models which is based on the mixture of
anomaly and gauge mediations.  In particular, the gluino CEDM
is induced by the physical phase of the gluino mass mass and the $CP$-violating
couplings of gluino, and then it generates the $CP$-violating Weinberg
operator.

We estimated the effect of the gluino CEDM in the extension of
high-scale SUSY scenarios, and we showed the nucleon EDMs induced by
the gluino CEDM and the Barr-Zee contributions. The dominant
contribution to the nucleon EDMs within the MSSM comes from the Barr-Zee contribution of
chargino/neutralino loops at two-loop level, while one-loop diagrams
of the messenger particles generate the gluino CEDM. We revealed that
the gluino CEDM may affect on the prediction of EDMs of nucleons in
the high-scale SUSY models, especially in the cases of the light
messengers or heavy higgsino in comparison with $M_S$.  We do not
determine the total EDMs for nucleons since we still have large ambiguities
in the naive dimensional analysis for the Weinberg operator, including
the sign. If it is determined precisely, the proton and neutron EDMs
may be found to behave differently in the extended models, and thus,
it would be more important to detect nucleon EDMs and electron EDM or
the ratios of them.

\section*{Acknowledgments}

We thank Dr. Natsumi Nagata for useful discussion.
This work is supported by Grant-in-Aid for Scientific research from
the Ministry of Education, Science, Sports, and Culture (MEXT), Japan,
No. 23104011 (J.H.). The work of J.H. is also supported by World
Premier International Research Center Initiative (WPI Initiative),
MEXT, Japan. The work of D.K. is supported by Grant-in-Aid for Japan Society
for the Promotion of Science (JSPS) Fellows (No. 26004521).
\newpage
\section*{Appendix}
\appendix

\section{Renormalization Group Equations\label{app:RGE}}

The (flavor-diagonal) $CP$-violating operators below the electroweak scale are given up to dimension-six operators as defined in \cref{eq:CP_viol}.
For considering the RGE evolutions of these operators, 
it is convenient to define the quark (C)EDM operators $\mathcal{O}^q_i ~ (i = 1,2)$ as 
\begin{equation}
\mathcal{O}^q_1 = -\frac{i}{2}m_qe e_q\ovl{q}(F\cdot \si)\gamma_5q\;, ~~~
\mathcal{O}^q_2 = -\frac{i}{2}m_q\ovl{q}g_3(G\cdot \si)\gamma_5q\;.
\end{equation}
$e_q$ denotes the electric charge of the quark $q$ and $g_3$ is the QCD coupling constant. 
The Weinberg operator is also given by $\mathcal{O}_W$ as in 
Eq.~(\ref{eq:weinberg_ceff}). 
Then, by using these operators, the effective Lagrangian is rewritten as follows:
\eqs{
\mathcal{L}_{\cancel{\mathrm{CP}}} & = \sum_{i=1}^2 \sum_{q=u,d,s} C^q_i \mathcal{O}^q_i + C_W \mathcal{O}_W.
}
The relation between the coefficients in the Lagrangian in \cref{eq:CP_viol} and the Wilson coefficients is given by:
\eqs{
d_q = m_q e e_q C^q_1\;, ~~~~~
\widetilde{d}_q = m_q C^q_2\;, ~~~~~
w = -g_3 C_W.
}
The RGEs for the Wilson coefficients at the leading order are given as \cite{Degrassi:2005zd}
\eqs{
\pdiff{1}{}{\ln\mu} \mathbf{C} = \mathbf{C} \cdot \mathbf{\Gamma}, \label{eq:RGE_C}
}
where 
\eqs{
\mathbf{C} = (C_1^q, C_2^q, C_W), ~~~~
\mathbf{\Gamma} = \frac{g_3^2}{16\pi^2} \left( 
\begin{array}{ccc}
8C_F & 0 & 0 \\
8C_F & 16C_F-4N_C & 0 \\
0 & 2N_C & N_C+2N_f - b_0
\end{array}
\right).
}
Here, $N_C(=3)$ and $N_f$ are the number of colors and quark flavors, respectively. 
$C_F=(N_C^2-1)/(2N_C)$ is the Casimir invariant and $b_0=-11N_C/3 + 2N_f/3 $ is the coefficient of the one-loop beta function for $g_3$.

\section{QCD Sum Rules for Nucleon EDMs \label{app:QCDsum}}

In order to get predictions for the nucleon and electron EDMs, 
we have to estimate the contribution to them from the parton-level interactions.
In Ref.~\cite{Hisano:2012sc}, the neutron EDM $d_n$ is related to the quark EDMs and CEDMs by using the QCD sum rules at the renormalization scale $\mu = 1~{\rm GeV}$.
Similarly, the proton EDM $d_p$ is also associated with the quark (C)EDMs at the renormalization scale $\mu = 1~{\rm GeV}$.
The nucleon EDMs are given by using the QCD sum rules as 
\eqs{
d_N = \frac{- c_0 m_N^3 \ave{\ovl qq}}{\lam_N^2} \Theta_N, ~~~~
(N = p, n).
}
Here, $c_0 = 0.234$, $\ave{\ovl q q} =-m_\pi^2f_\pi^2/(m_u+m_d) =- (0.262~{\rm GeV})^3$ is the quark condensate, 
and $\lam_N$ relates the interpolation fields with the proton and neutron fields.
$m_N$ denotes the mass of nucleon $N$.
$\Theta_N$ is calculated through the operator product expansions (OPE) for the correlator of interpolation fields and are just the coefficients proportional to $\ave{\ovl q q}$.
Without the Peccei-Quinn (PQ) mechanism \cite{Peccei:1977hh} for the strong $CP$ problem, we obtain
\eqs{
\Theta_p & 
= (4 e_u m_u \rho_u - e_d m_d \rho_d) \chi \ovl\theta + (4 d_u - d_d ) + \left( \kappa - \frac12 \xi \right) \left( 4 e_u \widetilde d_u - e_d \widetilde d_d \right), \\
\Theta_n & 
= (4 e_d m_d \rho_d - e_u m_u \rho_u) \chi \ovl\theta + (4 d_d - d_u ) + \left( \kappa - \frac12 \xi \right) \left( 4 e_d \widetilde d_d - e_u \widetilde d_u \right). \\
}
Here, $e_q$, $m_q$, and $d_q$ ($\widetilde d_q$)
 denote the electric charge for quark $q$, the mass of quark $q$, and the (C)EDM for quark $q$, respectively. $\chi, \kappa$, and $\xi$ are parameters which relate the quark condensates on the electromagnetic background with $\ave{\ovl q q}$; 
\eqs{
\ave{\ovl q \si_{\mu\nu} q}_F & = e_q \chi F_{\mu\nu} \ave{\ovl q q}, \\
g_s \ave{\ovl q G^A_{\mu\nu} T^A q}_F & = e_q \kappa F_{\mu\nu} \ave{\ovl q q}, \\
2 g_s \ave{\ovl q \widetilde G^A_{\mu\nu} T^A q}_F & = i e_q \xi F_{\mu\nu} \ave{\ovl q q}, \\
}
where $\ave{\cdots}_F$ denotes the vacuum expectation value on the electromagnetic background.
These parameters are estimated by Refs.~\cite{Belyaev:1982sa,Kogan:1991en}, and then the values are given by
$\chi = - 5.7~{\rm GeV}^{-2}, ~ \xi = -0.74$, and $\kappa = -0.34$.
$\rho_u$ and $\rho_d$ are defined as
\eqs{
\rho_u & 
= \frac{m_\ast}{m_u} \left\{ 1 + \frac{m_0^2}{2 \ovl \theta} \left[ \frac{\widetilde d_u - \widetilde d_d}{m_d} + \frac{\widetilde d_u - \widetilde d_s}{m_s} \right] \right\}, \\
\rho_d & 
= \frac{m_\ast}{m_d} \left\{ 1 + \frac{m_0^2}{2 \ovl \theta} \left[ \frac{\widetilde d_d - \widetilde d_u}{m_u} + \frac{\widetilde d_d - \widetilde d_s}{m_s} \right] \right\}.
}
The parameter $m_0^2$ which is associated with the VEV of $\ovl q (G \cdot \si) q$ is estimated by Belyaev and Ioffe \cite{Belyaev:1982sa}: $m_0^2 = 0.8~{\rm GeV}^2$. $m_\ast$ is the reduced quark mass defined as $m_\ast^{-1} = m_u^{-1} + m_d^{-1} + m_s^{-1}$.
The two-loop evolution of $\lam_N$ is given as follows:
\eqs{
\lam_N(1~{\rm GeV}) & = 
\left( \frac{\alpha_s(1~{\rm GeV})}{\alpha_s(m_c)} \right)^{-\frac29}\left( \frac{\alpha_s(m_c)}{\alpha_s(2~{\rm GeV})} \right)^{-\frac{6}{25}} \\
& \times \left( \frac{\alpha_s(1~{\rm GeV})+\frac{9\pi}{16}}{\alpha_s(m_c)+\frac{9\pi}{16}} \right)^{\frac29-\frac{41}{64}}
\left( \frac{\alpha_s(m_c)+\frac{100\pi}{154}}{\alpha_s(2~{\rm GeV})+\frac{100\pi}{154}} \right)^{\frac{6}{25}-\frac{117}{154}}
\lam_N(2~{\rm GeV}) \\
& = - 0.0439~{\rm GeV}^3,
}
for $ \lam_N(2~{\rm GeV}) = - 0.0480~{\rm GeV}^3$. 
Thus, we obtain the relation between the nucleon EDMs and the quark (C)EDMs as follows:
\eqs{
d_p & = - 1.2 \times 10^{-16}[e~{\rm cm}] \ovl\theta 
+ 0.78 d_u- 0.20d_d
+ e( -0.28 \widetilde d_u + 0.28 \widetilde d_d + 0.021 \widetilde d_s), \\
d_n & = 8.2 \times 10^{-17}[e~{\rm cm}] \ovl\theta 
- 0.12 d_u + 0.78 d_d
+ e( -0.30 \widetilde d_u + 0.30 \widetilde d_d - 0.014 \widetilde d_s). \\
}

Even if the PQ mechanism works, the theta parameter is induced as
\eqs{
\ovl\theta_{\rm ind} = \frac{m_0^2}{2} \sum_q \frac{\widetilde d_q}{m_q},
}
and then, we find the OPE coefficients $\Theta_N$ as 
\eqs{
\Theta_p & 
= (4 d_u - d_d ) + \left( \kappa - \frac12 \xi + \frac{m_0^2}{2} \chi \right) \left( 4 e_u \widetilde d_u - e_d \widetilde d_d \right), \\
\Theta_n & 
= (4 d_d - d_u ) + \left( \kappa - \frac12 \xi + \frac{m_0^2}{2} \chi \right) \left( 4 e_d \widetilde d_d - e_u \widetilde d_u \right). \\
}
Thus, we find the relation between the nucleon EDMs and the quark (C)EDMs under the PQ symmetry as follows:
\eqs{
d_p^{\rm PQ} & = 
 0.78 d_u - 0.20d_d
+ e( -1.2 \widetilde d_u - 0.15 \widetilde d_d ), \\
d_n^{\rm PQ} & = 
- 0.20 d_u + 0.78d_d
+ e( 0.29 \widetilde d_u + 0.59 \widetilde d_d). \\
}
\newpage
\bibliography{ref}
\end{document}